\documentclass[reprint,noshowpacs,noshowkeys,prd,balancelastpage,nofootinbib]{revtex4}
\usepackage{amsfonts}
\usepackage{mdframed}
\usepackage{amssymb}
\usepackage{footnote}
\usepackage{amsmath}
\usepackage{graphicx}
\usepackage{float}
\usepackage[font={footnotesize,it}]{caption}
\usepackage[utf8]{inputenc}
\usepackage{natbib}
\usepackage{tikz}
\usepackage{tikz-3dplot}
\usepackage[colorlinks=true,
            linkcolor=purple,
            urlcolor=purple,
            citecolor=blue]{hyperref}

\begin{document}

\title{Quantum particle on the surface of a spherocylindrical capsule}
\author{Elham Poorkahnooji}
\email{20620008@emu.edu.tr}
\author{S. Habib Mazharimousavi}
\email{habib.mazhari@emu.edu.tr}
\affiliation{Department of Physics, Faculty of Arts and Sciences, Eastern Mediterranean
University, Famagusta, North Cyprus via Mersin 10, T\"{u}rkiye}
\date{\today }

\begin{abstract}
A spinless nonrelativistic quantum particle on the curved surface of a
homogeneous spherocylindrical capsule is considered. We apply Costa's
formalism to solve the Schr\"{o}dinger equation with only a confined
potential forcing the particle to remain on the surface and be free to move.
It is shown that while a quantum particle with zero tangential/local energy
can exist on the surface of a spherical shell with an arbitrary radius, it
exists on a spherocylinder capsule only with a quantized length-to-radius
ratio. In other words, if and only if the length-to-radius ratio of the
capsule is an even multiplication of $\pi $, the wave function on the
surface interferes with itself constructively such that the wave function
survives. This hypothetical phenomenon may lead to applications in nanoscale
measurements.
\end{abstract}

\keywords{Curved surface; Quantum particle; Bound state; }
\maketitle

\section{Introduction}

Quantum particles confined on curved surfaces are a growing
multidisciplinary subject with wide applications in nanotechnology \cite%
{NT1,NT2}, for instance, condensed matter physics \cite{CMP}, material
science, and the study of electronic properties of nanostructured materials
such as graphene \cite{Gra,Gra2,Gra3,Gra4}, carbon nanotube \cite%
{Car,Car2,Car3}, and Phosphorant \cite{Pho}. The formalism was developed
initially by De Witt in 1957 \cite{Witt} which was suffering from the
so-called \textit{ordering-ambiguity}. A different approach was attempted by
Jensen and Koppe in 1971 \cite{JK} which showed the necessity of considering
the effects of geometry and probably topology in the formation of a
geometrical potential. In 1981, eventually, Costa in \cite{Costa} (see also 
\cite{Costa2}) analytically obtained the Schr\"{o}dinger equation for a
quantum particle confined to a 2-dimensional curved surface embedded in the
3-dimensional Euclidean space. Following \cite{Costa,Costa2} in 2008,
Ferrari and Cuoghi extended the formalism to the Schr\"{o}dinger equation
for a particle on a curved surface in an electric and magnetic field \cite%
{EM}. Moreover, another important extension of Costa's work has been made
recently by Liang and Lai in \cite{Lai} where the inhomogeneous confinement
- as a result of the variable thicknesses of the curved surface - has been
worked out. 

These formalisms i.e., \cite{Costa,Costa2,EM} have been widely employed in
the literature. For instance, very recently Meschede et al studied the
quantum scattering of a nonrelativistic spinless quantum particle confined
on a curved surface from the geometric potential due to the localized
curvatures of the surface \cite{Mes}. In \cite{Lai2} the \textit{%
geometry-induced wave-function collapse} has been studied which is in
analogy with the collapse of a quantum system in 3-dimensions undergoing a
potential in the form of an inverse squared distance. Although, according to 
\cite{Lai2}, such a potential geometrically forms a truncated conic surface
instead of a Coulomb impurity. In \cite{PDM}, the effects of the
position-dependent mass of an electron confined on a bilayer graphene
catenoid bridge on its electronic properties are investigated. Gravesen et
al in \cite{MP} studied the quantum particle on a generic surface of
revolution and in an interesting example, they obtained the eigenvalues and
eigenfunctions of the quantum particle confined on a finite cylindrical
surface by applying two different methods. First, they employed Costa's
formalism where the thickness of the finite cylinder was going to zero. Then
in \cite{MP} the authors assumed the quantum particle was confined in a
cylindrical hollow where the Cylindrical surface is considered thick such
that after they obtained the eigenvalues and eigenfunctions by solving the
Schr\"{o}dinger equation in 3 dimensions they calculated the limits when the
thickness of the cylindrical hollow goes to zero. As was reported in \cite%
{MP}, the results of the two methods agree which confirms the feasibility of
Costa's formalism. 

Considering all these efforts, in this current research we study a
nonrelativistic spinless quantum particle on a particular compact surface
called the "\textit{spherocylindrical capsule}". Although confined quantum
particles on some nontrivial surfaces have been studied in the literature
(see for example \cite{S1,S2,S3,S4,S5,S6,S7,S8,S9}), we aim to report an
important feature of this special geometry with a quantum particle of zero
local energy confined on its surface. Here is the organization of this paper. In Sec. II we revisit
Costa's formalism briefly. In Sec. III and IV, we introduce the geometry of
the sphere and spherocylindrical capsule and solve the corresponding Schr%
\"{o}dinger equations. We conclude our paper by giving an interpretation and
possible applications in Sec. V.

\section{Schr\"{o}dinger equation on a curved surface: a brief review}

Consider a non-relativistic quantum particle constrained to remain on a
curved surface $\mathcal{S}.$ Herein, the curved surface $\mathcal{S}$ is
the map $\mathbf{f}\left( u,v\right) :%
\mathbb{R}
^{2}\supset U\rightarrow \mathcal{S}\subset 
\mathbb{R}
^{3}$ where the finite or infinite plane $U=\left\{ \left( u,v\right) \in 
\mathbb{R}
^{2}|u_{1}<u<u_{2},v_{1}<v<v_{2}\right\} $ is mapped into the surface $%
\mathcal{S}\subset 
\mathbb{R}
^{3}.$ This quantum particle has to satisfy the Schr\"{o}dinger equation 
\begin{equation}
-\frac{\hbar ^{2}}{2M}\frac{1}{\sqrt{g}}\partial _{\mu }\left( \sqrt{g}%
g^{\mu \nu }\partial _{\nu }\right) \psi \left( x^{\mu },t\right)
+V_{c}\left( x^{3}\right) \psi \left( x^{\mu },t\right) =i\hbar \partial
_{t}\psi \left( x^{\mu },t\right) ,  \label{1}
\end{equation}%
in which $M$ is the mass of the particle, $g_{\mu \nu }$ is the metric
tensor of the Euclidean space $%
\mathbb{R}
^{3}$ with $g=\det g_{\mu \nu }$, $\mu \in \left\{ 1,2,3\right\} $ such that 
$x^{1}=u,$ $x^{2}=v$ are used in a two dimensional local coordinates system
on the curved surface and $x^{3}=w$ is the distance of any point in the
space from the surface $\mathcal{S}$. Furthermore, $V_{c}\left( x^{3}\right) 
$ is the constraint potential such that for a homogeneous and uniform
thickness surface it is zero when $x^{3}\subset \left( -\frac{\varepsilon }{2%
},\frac{\varepsilon }{2}\right) $ with $\varepsilon \rightarrow 0$ the
uniform thickness of the surface and infinity elsewhere. In other words, the
particle experiences a one-dimensional infinite well potential across the
thickness of the surface. Therefore, one writes%
\begin{equation}
\psi \left( x^{\mu },t\right) =e^{-iEt/\hbar }\psi ^{\left( N\right) }\left(
w\right) \psi ^{\left( T\right) }\left( u,v\right) ,  \label{2}
\end{equation}%
where the subindices $N$ and $T$ stand for the normal and tangent wave
function, respectively. Introducing (\ref{2}) in (\ref{1}) and separating
the variables yields the following separated equations%
\begin{equation}
-\frac{\hbar ^{2}}{2M}\psi ^{\left( N\right) \prime \prime }\left( w\right)
+V_{c}\left( w\right) \psi ^{\left( N\right) }\left( w\right) =E^{\left(
N\right) }\psi ^{\left( N\right) }\left( w\right) ,  \label{3}
\end{equation}%
and 
\begin{equation}
-\frac{\hbar ^{2}}{2M}\frac{1}{\sqrt{h}}\partial _{a}\left( \sqrt{h}%
h^{ab}\partial _{b}\right) \psi ^{\left( T\right) }\left( x^{c}\right)
+V_{g}\left( x^{c}\right) \psi ^{\left( T\right) }\left( x^{c}\right)
=E^{\left( T\right) }\psi ^{\left( T\right) }\left( x^{c}\right) ,  \label{4}
\end{equation}%
where $a,b,c\in \left\{ 1,2\right\} $, $h_{ab}$ is the induced metric tensor
on the curved surface $\mathcal{S}$ with $h=\det h_{ab}$, $E=E^{\left(
N\right) }+E^{\left( T\right) }$ is the total energy of the particle with $%
E^{\left( N\right) }$ the normal/global energy due to the confinement force
and $E^{\left( T\right) }$ is the tangent/local energy due to geometrical
potential given by 
\begin{equation}
V_{g}\left( x^{c}\right) =-\frac{\hbar ^{2}}{2M}\left( \frac{1}{4}%
K^{2}-K_{G}\right) ,  \label{5}
\end{equation}%
with $K$ and $K_{G}$ the total and the Gaussian curvature of $\mathcal{S}$.
For a detailed proof, we refer to \cite{S9}. The normal Schr\"{o}dinger
equation can be solved to get%
\begin{equation}
\psi _{n}^{\left( N\right) }\left( w\right) =\sqrt{\frac{\varepsilon }{2}}%
\sin \left( \frac{n\pi }{\varepsilon }w\right) ,  \label{6}
\end{equation}%
and%
\begin{equation}
E_{n}^{\left( N\right) }=\frac{n^{2}\pi ^{2}\hbar ^{2}}{2M\varepsilon ^{2}},%
\text{ }n=1,2,3,...\text{.}  \label{7}
\end{equation}%
In solving (\ref{3}), we assumed the surface to be thin with the thickness $%
\varepsilon $ such that the constrained potential simplifies to 
\begin{equation}
V_{c}\left( w\right) =\left\{ 
\begin{array}{cc}
0, & 0<w<\varepsilon  \\ 
\infty , & elsewhere%
\end{array}%
\right. .  \label{8}
\end{equation}%
While the normal part of the wave function of the constrained quantum
particle is trivially given by (\ref{6}) for all homogenous surfaces, the
tangent equation (\ref{4}) strongly depends on the geometrical potential (%
\ref{5}) which is the effect of the curvatures of the surface. For surfaces
that are not intrinsically curved, the Gaussian curvature is zero, and the
geometric potential becomes negative definite. For instance, for a flat
plane, both curvatures are zero and consequently $V_{G}\left( w\right) =0.$
A circular cylinder of radius $R$ is not intrinsically curved and therefore
one finds $K_{G}=0$ and $K=\frac{1}{R}$ which amount $V_{G}\left( w\right) =-%
\frac{\hbar ^{2}}{2M}\frac{1}{4R^{2}}.$ Unlike a cylinder, a spherical
surface of radius $R$ is intrinsically curved such that both $K$ and $K_{G}$
are nonzero and given by $K=\frac{2}{R}$ and $K_{G}=\frac{1}{R^{2}}$ that
result in a zero geometrical potential i.e., $V_{G}\left( w\right) =0$.

\section{Zero-tangent/local energy particle on a sphere}

Let us assume a quantum particle with a zero tangent or local energy located
on a spherical shell of radius $R$ and thickness $\varepsilon .$ Having the
tangent energy to be zero implies that the total energy of the particle is 
\begin{equation}
E_{n}=E_{n}^{\left( N\right) }=\frac{n^{2}\pi ^{2}\hbar ^{2}}{2M\varepsilon
^{2}},\text{ }n=1,2,3,...\text{.}  \label{9}
\end{equation}%
The tangent Schr\"{o}dinger equation (\ref{4}) becomes%
\begin{equation}
-\frac{\hbar ^{2}}{2M}\frac{1}{\sqrt{h}}\partial _{a}\left( \sqrt{h}%
h^{ab}\partial _{b}\right) \psi ^{\left( T\right) }\left( x^{c}\right) =0,
\label{10}
\end{equation}%
where $u=\theta $ and $v=\varphi $ are the local coordinates with $\theta $
and $\varphi $ being the standard polar and azimuthal angles in the
spherical coordinates system. Furthermore, the induced metric tensor of the
shell is found to be 
\begin{equation}
h_{ab}=\left( 
\begin{array}{cc}
R^{2} & 0 \\ 
0 & R^{2}\sin ^{2}\theta 
\end{array}%
\right) ,  \label{11}
\end{equation}%
with $\sqrt{h}=R^{2}\sin \theta .$ Equation (\ref{10}) hence becomes%
\begin{equation}
\frac{1}{\sin \theta }\partial _{\theta }\left( \sin \theta \partial
_{\theta }\right) \psi ^{\left( T\right) }\left( \theta ,\varphi \right) +%
\frac{1}{\sin ^{2}\theta }\partial _{\varphi }\partial _{\varphi }\psi
^{\left( T\right) }\left( \theta ,\varphi \right) =0.  \label{12}
\end{equation}%
Next, we introduce 
\begin{equation}
\psi ^{\left( T\right) }\left( \theta ,\varphi \right) =\Theta \left( \theta
\right) e^{im\varphi },  \label{13}
\end{equation}%
in (\ref{12}) to get%
\begin{equation}
\frac{1}{\sin \theta }\partial _{\theta }\left( \sin \theta \partial
_{\theta }\right) \Theta \left( \theta \right) -\frac{m^{2}}{\sin ^{2}\theta 
}\Theta \left( \theta \right) =0.  \label{14}
\end{equation}%
With $m$ as an integer number due to the azimuthal symmetry, the only
possible solution that remains finite at $\theta =0$ and $\theta =\pi $ is a
constant corresponding to $m=0$. Therefore one writes%
\begin{equation}
\psi _{m=0}^{\left( T\right) }\left( \theta ,\varphi \right) =C,  \label{15}
\end{equation}%
where $C$ is an integration constant. To normalize the solution we write 
\begin{equation}
\int \int\nolimits_{S}\left\vert \psi ^{\left( T\right) }\left( \theta
,\varphi \right) \right\vert ^{2}R^{2}\sin \theta d\theta d\varphi =1,
\label{16}
\end{equation}%
which yields%
\begin{equation}
C=\frac{1}{\sqrt{4\pi R^{2}}},  \label{17}
\end{equation}%
and therefore the full solution becomes 
\begin{equation}
\psi _{n}\left( \theta ,\varphi ,w,t\right) =\sqrt{\frac{\varepsilon }{8\pi
R^{2}}}\sin \left( \frac{n\pi }{\varepsilon }w\right) e^{-iE_{n}t/\hbar },%
\text{ }n=1,2,3,...,  \label{18}
\end{equation}%
with energy given by (\ref{9}). Before we move on it should be noted that
Eq. (\ref{10}) is the angular part of the 3-dimensional Laplace equation
whose solutions are known to be the spherical harmonics i.e., $\psi ^{\left(
T\right) }\left( \theta ,\varphi \right) =Y_{\ell ,m}\left( \theta ,\varphi
\right) $ where $\ell =0,1,2,...$ and $\left\vert m\right\vert \leq \ell .$
The normalized wave function, therefore, is proportion to $Y_{0,0}\left(
\theta ,\varphi \right) =\frac{1}{\sqrt{4\pi }}$.

\section{Zero-tangent energy particle on a spherocylindrical capsule}

\begin{figure}[tbp]
\includegraphics[width=100mm,scale=1]{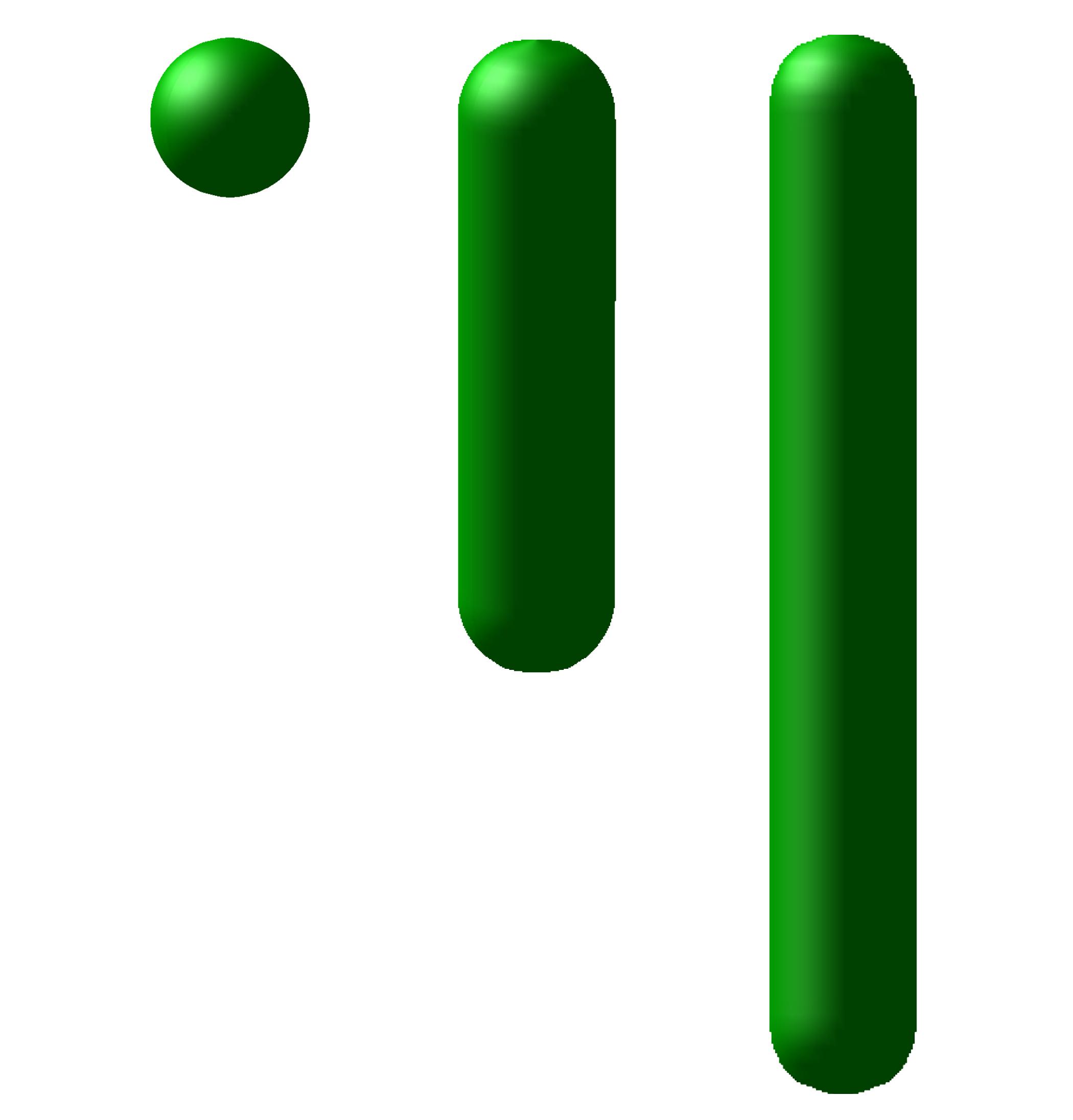}
\caption{Plot of the spherocylindrical capsule with the length-to-radius
ratio equal to $0$, $2\protect\pi $ and $4\protect\pi $ in actual scales.}
\label{F1}
\end{figure}
In this section, we solve the tangential Schr\"{o}dinger equation (\ref{4})
on the surface of the so-called spherocylindrical capsule. A
spherocylindrical capsule is a compact surface consisting of a circular
finite cylinder of radius $R$ and length $L$ which is closed on both ends by
a hemisphere of radius $R$ as is depicted in Fig. \ref{F1}. To parametrize
the entire surface homogeneously, we introduce the map 
\begin{equation}
\mathbf{f}\left( z,\varphi \right) =z\hat{k}+a\left( z\right) \left( \hat{%
\imath}\cos \varphi +\hat{\jmath}\sin \varphi \right) ,  \label{19}
\end{equation}%
in which $-R-\frac{L}{2}<z<\frac{L}{2}+R$ is the axis of the capsule and $%
0<\varphi <2\pi $ is the usual azimuthal angle. Furthermore, $a\left(
z\right) $ is the variable radius of the cross-section of the capsule at any
given $z$ expressed by%
\begin{equation}
a\left( z\right) =\left\{ 
\begin{array}{lr}
\sqrt{R^{2}-\left( z+\frac{L}{2}\right) ^{2}} & -R-\frac{L}{2}\leq z\leq -%
\frac{L}{2} \\ 
R & -\frac{L}{2}\leq z\leq \frac{L}{2} \\ 
\sqrt{R^{2}-\left( z-\frac{L}{2}\right) ^{2}} & \frac{L}{2}\leq z\leq \frac{L%
}{2}+R%
\end{array}%
\right. ,  \label{20}
\end{equation}%
and $\hat{\imath},\hat{\jmath}$ and $\hat{k}$ are the usual constant unit
vectors. Having the surface patch defined in (\ref{19}) and (\ref{20}), we
proceed to obtain the first fundamental form which is obtained to be%
\begin{equation}
h_{ab}=\left( 
\begin{array}{cc}
\mathbf{f}_{,z}.\mathbf{f}_{,z} & \mathbf{f}_{,z}.\mathbf{f}_{,\varphi } \\ 
\mathbf{f}_{,\varphi }.\mathbf{f}_{,z} & \mathbf{f}_{,\varphi }.\mathbf{f}%
_{,\varphi }%
\end{array}%
\right) =\left( 
\begin{array}{cc}
1+a^{\prime }\left( z\right) ^{2} & 0 \\ 
0 & a\left( z\right) ^{2}%
\end{array}%
\right) ,  \label{21}
\end{equation}%
where a prime stands for the derivative with respect to $z.$ Moreover, we
introduce the normal unit vector to the surface (outward direction) 
\begin{equation}
\hat{N}=-\frac{\mathbf{f}_{,z}\times \mathbf{f}_{,\varphi }}{\left\Vert 
\mathbf{f}_{,z}\times \mathbf{f}_{,\varphi }\right\Vert }=\frac{\hat{\imath}%
\cos \varphi +\hat{\jmath}\sin \varphi -a^{\prime }\hat{k}}{\sqrt{%
1+a^{\prime }\left( z\right) ^{2}}},  \label{22}
\end{equation}%
upon which we calculate the second fundamental form i.e.,%
\begin{equation}
\kappa _{ab}=\left( 
\begin{array}{cc}
\mathbf{f}_{,zz}.\hat{N} & \mathbf{f}_{,z\varphi }.\hat{N} \\ 
\mathbf{f}_{,\varphi z}.\hat{N} & \mathbf{f}_{,\varphi \varphi }.\hat{N}%
\end{array}%
\right) =\frac{1}{\sqrt{1+a^{\prime }\left( z\right) ^{2}}}\left( 
\begin{array}{cc}
a^{\prime \prime } & 0 \\ 
0 & -a%
\end{array}%
\right) .  \label{23}
\end{equation}%
Finally, we calculate the mixed extrinsic curvature tensor%
\begin{equation}
\kappa _{a}^{b}=h^{bc}\kappa _{ac}=\frac{1}{\sqrt{1+a^{\prime }\left(
z\right) ^{2}}}\left( 
\begin{array}{cc}
\frac{a^{\prime \prime }}{1+a^{\prime }\left( z\right) ^{2}} & 0 \\ 
0 & -\frac{1}{a}%
\end{array}%
\right) ,  \label{24}
\end{equation}%
from which we calculate the total and Gaussian curvatures i.e.,%
\begin{equation}
K=\text{Tr}\left( \kappa _{a}^{b}\right) =\frac{1}{\sqrt{1+a^{\prime }\left(
z\right) ^{2}}}\left( \frac{a^{\prime \prime }}{1+a^{\prime }\left( z\right)
^{2}}-\frac{1}{a}\right) ,  \label{25}
\end{equation}%
and%
\begin{equation}
K_{G}=\det \left( \kappa _{a}^{b}\right) =\frac{-a^{\prime \prime }}{a\left(
1+a^{\prime }\left( z\right) ^{2}\right) ^{2}}.  \label{26}
\end{equation}%
Having $K$ and $K_{G}$ calculated in (\ref{25}) and (\ref{26}), the
geometric potential (\ref{5}) becomes%
\begin{equation}
V_{G}\left( z,\varphi \right) =\left\{ 
\begin{array}{lr}
0 & -R-\frac{L}{2}\leq z\leq -\frac{L}{2} \\ 
-\frac{\hbar ^{2}}{8R^{2}M} & -\frac{L}{2}\leq z\leq \frac{L}{2} \\ 
0 & \frac{L}{2}\leq z\leq \frac{L}{2}+R%
\end{array}%
\right. ,  \label{27}
\end{equation}%
as was expected. Next, we introduce the tangential/local wave function in
the form%
\begin{equation}
\psi ^{\left( T\right) }\left( z,\varphi \right) =e^{im\varphi }Y\left(
z\right) ,  \label{28}
\end{equation}%
such that (\ref{4}) with zero-tangential/local energy becomes%
\begin{equation}
-\frac{\hbar ^{2}}{2M}\frac{1}{\sqrt{1+a^{\prime }\left( z\right) ^{2}}}%
\frac{d}{dz}\left( \frac{1}{\sqrt{1+a^{\prime }\left( z\right) ^{2}}}\frac{d%
}{dz}\right) Y\left( z\right) +\frac{\hbar ^{2}m^{2}}{2Ma\left( z\right) ^{2}%
}Y\left( z\right) +V_{G}Y\left( z\right) =0,  \label{29}
\end{equation}%
in which $m$ is an integer number due to the azimuthal symmetry of the
capsule. Knowing that on the surface%
\begin{equation}
\sqrt{1+a^{\prime }\left( z\right) ^{2}}=\frac{R}{a\left( z\right) },
\label{30}
\end{equation}%
the explicit form of (\ref{29}) is given by%
\begin{equation}
-\frac{\hbar ^{2}}{2M}\frac{a\left( z\right) }{R}\frac{d}{dz}\left( \frac{%
a\left( z\right) }{R}\frac{d}{dz}\right) Y\left( z\right) +\frac{\hbar
^{2}m^{2}}{2Ma\left( z\right) ^{2}}Y\left( z\right) +V_{G}Y\left( z\right)
=0.  \label{31}
\end{equation}%
In terms of different regions on the capsule, (\ref{31}) reads%
\begin{equation}
\left\{ 
\begin{array}{lr}
-\frac{\hbar ^{2}}{2M}\frac{\sqrt{R^{2}-\left( z+\frac{L}{2}\right) ^{2}}}{R}%
\frac{d}{dz}\left( \frac{\sqrt{R^{2}-\left( z+\frac{L}{2}\right) ^{2}}}{R}%
\frac{d}{dz}\right) Y_{I}\left( z\right) +\frac{\hbar ^{2}m^{2}}{2M\left(
R^{2}-\left( z+\frac{L}{2}\right) ^{2}\right) }Y_{I}\left( z\right) =0 & -R-%
\frac{L}{2}\leq z\leq -\frac{L}{2} \\ 
-\frac{\hbar ^{2}}{2M}\frac{d^{2}}{dz^{2}}Y_{II}\left( z\right) +\frac{\hbar
^{2}}{2MR^{2}}\left( m^{2}-\frac{1}{4}\right) Y_{II}\left( z\right) =0 & -%
\frac{L}{2}\leq z\leq \frac{L}{2} \\ 
-\frac{\hbar ^{2}}{2M}\frac{\sqrt{R^{2}-\left( z-\frac{L}{2}\right) ^{2}}}{R}%
\frac{d}{dz}\left( \frac{\sqrt{R^{2}-\left( z-\frac{L}{2}\right) ^{2}}}{R}%
\frac{d}{dz}\right) Y_{III}\left( z\right) +\frac{\hbar ^{2}m^{2}}{2M\left(
R^{2}-\left( z-\frac{L}{2}\right) ^{2}\right) }Y_{III}\left( z\right) =0 & 
\frac{L}{2}\leq z\leq \frac{L}{2}+R%
\end{array}%
\right. ,  \label{32}
\end{equation}%
where we used the subindices I, II, and III for the left, middle, and right
sides of the capsule. By introducing $z=-\frac{L}{2}+qR$ and $L=\lambda R$
in which $q$ is the new variable and $\lambda =\frac{L}{R}$ being the
length-to-radius ratio, the first equation in (\ref{32}) simplifies to%
\begin{equation}
-\sqrt{1-q^{2}}\frac{d}{dq}\left( \sqrt{1-q^{2}}\frac{d}{dq}\right)
Y_{I}\left( q\right) +\frac{m^{2}}{1-q^{2}}Y_{I}\left( q\right) =0,
\label{33}
\end{equation}%
which is the same as Eq. (\ref{14}) after considering $\cos \theta =q$. With 
$-R-\frac{L}{2}\leq z\leq -\frac{L}{2}$, we find $-1\leq q\leq 0$ and the
only finite solution of (\ref{33}) in this interval corresponds to $m=0$ and 
\begin{equation}
Y_{I}\left( q\right) =C_{I},  \label{34}
\end{equation}%
in which $C_{I}$ is an integration constant. Similarly in the third region,
we introduce $z=\frac{L}{2}+qR$ and $L=\lambda R$ to get%
\begin{equation}
-\sqrt{1-q^{2}}\frac{d}{dz}\left( \sqrt{1-q^{2}}\frac{d}{dz}\right)
Y_{III}\left( z\right) +\frac{m^{2}}{1-q^{2}}Y_{III}\left( z\right) =0,
\label{35}
\end{equation}%
such that $0\leq q\leq 1$. With the same reason here also $m=0$ and the only
solution being finite in the domain of $q$ is 
\begin{equation}
Y_{III}\left( q\right) =C_{III}.  \label{36}
\end{equation}%
Finally, the tangential Schr\"{o}dinger equation in the middle region after
introducing $m=0$ due to the continuity condition of the wave function
becomes%
\begin{equation}
\frac{d^{2}}{dz^{2}}Y_{II}\left( z\right) +\frac{1}{4R^{2}}Y_{II}\left(
z\right) =0,  \label{37}
\end{equation}%
which admits a solution in the form of%
\begin{equation}
Y_{II}\left( z\right) =A\cos \left( \frac{z}{2R}\right) +B\sin \left( \frac{z%
}{2R}\right) ,  \label{38}
\end{equation}%
in which $A$ and $B$ are two integration constants. After we obtained the
solutions on all regions we have to apply the boundary conditions. In
particular since we set $m=0$ the azimuthal part of the wave function is
trivially continuous. The non-azimuthal part of the wave function i.e., $%
Y\left( z\right) $ and its first derivative i.e., $Y^{\prime }\left(
z\right) $ have to be continuous everywhere on the surface of the capsule. In
particular we apply these conditions at $z=-\frac{L}{2}$ and $z=\frac{L}{2}.$
The wave function and its first derivatives are given by%
\begin{equation}
Y\left( z\right) =\left\{ 
\begin{array}{lr}
C_{I} & -R-\frac{L}{2}\leq z\leq -\frac{L}{2} \\ 
A\cos \left( \frac{z}{2R}\right) +B\sin \left( \frac{z}{2R}\right)  & -\frac{%
L}{2}\leq z\leq \frac{L}{2} \\ 
C_{III} & \frac{L}{2}\leq z\leq \frac{L}{2}+R%
\end{array}%
\right. ,  \label{39}
\end{equation}%
and%
\begin{equation}
Y^{\prime }\left( z\right) =\left\{ 
\begin{array}{lr}
0 & -R-\frac{L}{2}\leq z\leq -\frac{L}{2} \\ 
-\frac{A}{2R}\sin \left( \frac{z}{2R}\right) +\frac{B}{2R}\cos \left( \frac{z%
}{2R}\right)  & -\frac{L}{2}\leq z\leq \frac{L}{2} \\ 
0 & \frac{L}{2}\leq z\leq \frac{L}{2}+R%
\end{array}%
\right. ,  \label{40}
\end{equation}%
respectively. The continuity of the wave function and its derivative at $z=-%
\frac{L}{2}$ and $z=\frac{L}{2}$ yield%
\begin{equation}
\text{the boundary conditions}\Longrightarrow \left\{ 
\begin{array}{l}
A\cos \left( \frac{L}{4R}\right) -B\sin \left( \frac{L}{4R}\right) =C_{I} \\ 
A\cos \left( \frac{L}{4R}\right) +B\sin \left( \frac{L}{4R}\right) =C_{III}
\\ 
A\sin \left( \frac{L}{4R}\right) +B\cos \left( \frac{L}{42R}\right) =0 \\ 
-A\sin \left( \frac{L}{4R}\right) +B\cos \left( \frac{L}{4R}\right) =0%
\end{array}%
\right. .  \label{41}
\end{equation}%
\begin{figure}[tbp]
\includegraphics[width=120mm,scale=1]{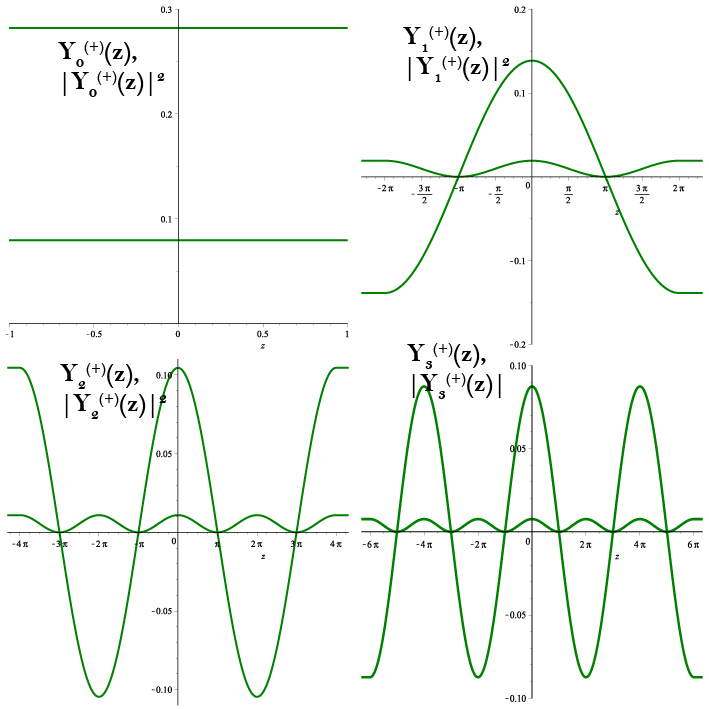}
\caption{Plots of the wavefunctions $Y_{k}^{\left( +\right) }$ and $%
\left\vert Y_{k}^{\left( +\right) }\right\vert ^{2}$ in terms of the axial
variable $z$ for the first four even states i.e., $k=0,1,2$ and $4$
corresponding to the length-to-radius ration $\protect\lambda _{k}^{\left(
+\right) }=\frac{L_{k}^{\left( +\right) }}{R}=4k\protect\pi .$ The upper
panel left plot corresponds to the spherical capsule where the probability
density is uniform over the surface of the sphere. This uniform probability
can be seen in the other three graphs only on the hemispheres in the left
and the right of the graph. The probability on the cylindrical part of the
capsule, however, is not uniform and instead is periodic.}
\label{F2}
\end{figure}
\begin{figure}[tbp]
\includegraphics[width=120mm,scale=1]{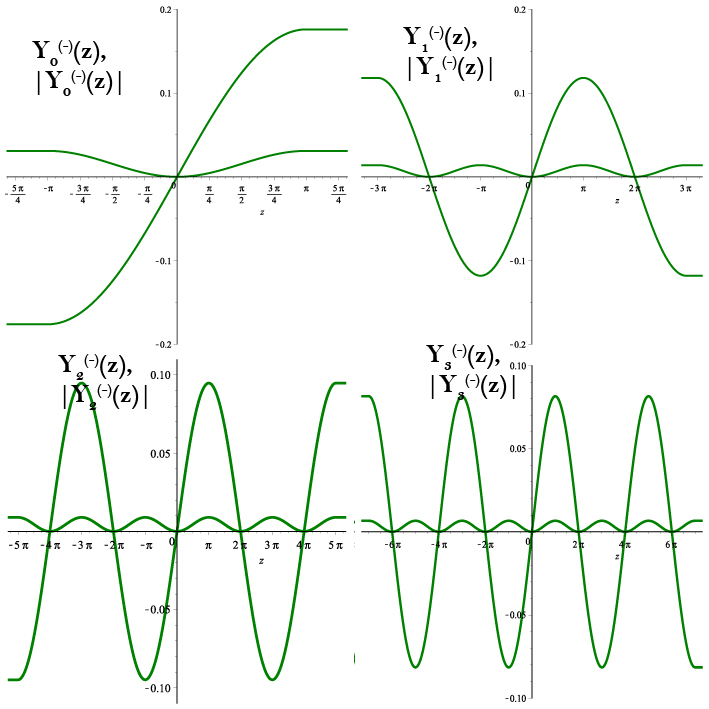}
\caption{Plots of the wavefunctions $Y_{k}^{\left( -\right) }$ and $%
\left\vert Y_{k}^{\left( -\right) }\right\vert ^{2}$ in terms of the axial
variable $z$ for the first four odd states i.e., $k=0,1,2$ and $3$
corresponding to the length-to-radius ration $\protect\lambda _{k}^{\left(
-\right) }=\frac{L_{k}^{\left( -\right) }}{R}=2\left( 2k+1\right) \protect%
\pi .$ Similar to the even states, here also the probability density on the
two hemispheres is uniform and on the cylindrical part of the capsule is
periodic.}
\label{F3}
\end{figure}
The last two equations admit nontrivial solutions for $A$ and $B$ if and
only if 
\begin{equation}
\sin \left( \frac{L}{4R}\right) \cos \left( \frac{L}{4R}\right) =0.
\label{42}
\end{equation}%
Hence, while with $\cos \left( \frac{L}{2R}\right) =0$ one gets $A=0$ and $%
B\neq 0$, with $\sin \left( \frac{L}{2R}\right) =0$ it yields $B=0$ and $%
A\neq 0.$ For the former case the first two equations imply $%
C_{III}=-C_{I}=B\sin \left( \frac{L}{2R}\right) $ and therefore the solution
becomes%
\begin{equation}
Y^{\left( -\right) }\left( z\right) =\left\{ 
\begin{array}{lr}
-B\sin \left( \frac{L}{4R}\right)  & -R-\frac{L}{2}\leq z\leq -\frac{L}{2}
\\ 
B\sin \left( \frac{z}{4R}\right)  & -\frac{L}{2}\leq z\leq \frac{L}{2} \\ 
B\sin \left( \frac{z}{4R}\right)  & \frac{L}{2}\leq z\leq \frac{L}{2}+R%
\end{array}%
\right. \text{ with }\frac{L}{4R}=\left( k+\frac{1}{2}\right) \pi .\text{ }
\label{43}
\end{equation}%
On the other hand for the latter case, the first two equations yield $%
C_{III}=C_{I}=A\cos \left( \frac{L}{4R}\right) $ and the wave function reads 
\begin{equation}
Y^{\left( +\right) }\left( z\right) =\left\{ 
\begin{array}{lr}
A\cos \left( \frac{L}{4R}\right)  & -R-\frac{L}{2}\leq z\leq -\frac{L}{2} \\ 
A\cos \left( \frac{z}{4R}\right)  & -\frac{L}{2}\leq z\leq \frac{L}{2} \\ 
A\cos \left( \frac{L}{4R}\right)  & \frac{L}{2}\leq z\leq \frac{L}{2}+R%
\end{array}%
\right. \text{ with }\frac{L}{4R}=k\pi .  \label{44}
\end{equation}%
Herein, the superindices $\pm $ stand for the even and odd solution and $%
k=0,1,2,3,...$. To normalize the wave function, we write 
\begin{equation}
\int_{0}^{2\pi }\int_{-R-\frac{L}{2}}^{R+\frac{L}{2}}\left\vert Y^{\left(
\pm \right) }\left( z\right) \right\vert ^{2}\sqrt{h}dzd\varphi =1,
\label{45}
\end{equation}%
in which 
\begin{equation}
h=\det h_{ab}=\left( 1+a^{\prime }\left( z\right) ^{2}\right) a\left(
z\right) ^{2}=R^{2},  \label{46}
\end{equation}%
in all regions. Explicitly (\ref{45}) for the even solution reads%
\begin{equation}
2\pi R\left\vert A\right\vert ^{2}\left( \int_{-R-\frac{L}{2}}^{-\frac{L}{2}%
}dz+\int_{-\frac{L}{2}}^{+\frac{L}{2}}\cos ^{2}\left( \frac{z}{2R}\right)
dz+\int_{\frac{L}{2}}^{\frac{L}{2}+R}dz\right) =1,  \label{47}
\end{equation}%
in which we used $\cos \left( \frac{L}{4R}\right) =\cos \left( k\pi \right)
=\left( -1\right) ^{k}.$ Hence, we obtain 
\begin{equation}
\left\vert A\right\vert ^{2}=\frac{1}{2\pi R\left( 2R+\frac{L}{2}\right) }=%
\frac{1}{4\pi R^{2}\left( 1+k\pi \right) }.  \label{48}
\end{equation}%
A similar calculation for the odd wave function yields%
\begin{equation}
\left\vert B\right\vert ^{2}=\frac{1}{4\pi R^{2}\left( 1+\left( k+\frac{1}{2}%
\right) \pi \right) }.  \label{49}
\end{equation}%
Finally, the normalized wave function is expressed by%
\begin{equation}
\psi _{n,k}^{\left( +\right) }\left( z,\varphi ,w,t\right) =\sqrt{\frac{%
\varepsilon }{8\pi R^{2}\left( 1+k\pi \right) }}\left\{ 
\begin{array}{lr}
\left( -1\right) ^{k}\sin \left( \frac{n\pi }{\varepsilon }w\right)
e^{-iE_{n}t/\hbar } & -R-\frac{L}{2}\leq z\leq -\frac{L}{2} \\ 
\cos \left( \frac{z}{2R}\right) \sin \left( \frac{n\pi }{\varepsilon }%
w\right) e^{-iE_{n}t/\hbar } & -\frac{L}{2}\leq z\leq \frac{L}{2} \\ 
\left( -1\right) ^{k}\sin \left( \frac{n\pi }{\varepsilon }w\right)
e^{-iE_{n}t/\hbar } & \frac{L}{2}\leq z\leq \frac{L}{2}+R%
\end{array}%
\right. ,  \label{50}
\end{equation}%
and%
\begin{equation}
\psi _{n,k}^{\left( -\right) }\left( z,\varphi ,w,t\right) =\sqrt{\frac{%
\varepsilon }{8\pi R^{2}\left( 1+\left( k+\frac{1}{2}\right) \pi \right) }}%
\left\{ 
\begin{array}{lr}
\left( -1\right) ^{k+1}\sin \left( \frac{n\pi }{\varepsilon }w\right)
e^{-iE_{n}t/\hbar } & -R-\frac{L}{2}\leq z\leq -\frac{L}{2} \\ 
\sin \left( \frac{z}{2R}\right) \sin \left( \frac{n\pi }{\varepsilon }%
w\right) e^{-iE_{n}t/\hbar } & -\frac{L}{2}\leq z\leq \frac{L}{2} \\ 
\left( -1\right) ^{k}\sin \left( \frac{n\pi }{\varepsilon }w\right)
e^{-iE_{n}t/\hbar } & \frac{L}{2}\leq z\leq \frac{L}{2}+R%
\end{array}%
\right. .  \label{51}
\end{equation}%
We note that, in both cases, the total energy is equal to the normal/global
energy i.e., $E_{n}=E_{n}^{\left( N\right) }=\frac{n^{2}\pi ^{2}\hbar ^{2}}{%
2M\varepsilon ^{2}},$ $n=1,2,3,...$ because the local energy was set to
zero. Furthermore, the wave functions (\ref{50}) and (\ref{51}) both exist
only if and only if the length-to-radius ratio $\lambda _{k}^{\left( \pm
\right) }=\frac{L}{R}$ is quantized. The first quantized length is $\lambda
_{0}^{\left( +\right) }=0$ which corresponds to the spherical shell. In
other words, the capsule is a sphere that has been studied separately in
the previous section. The second, third, and fourth quantized
length-to-radius ratio correspond to $\lambda _{0}^{\left( -\right) }=\pi $, 
$\lambda _{1}^{\left( +\right) }=2\pi $ and $\lambda _{1}^{\left( -\right)
}=3\pi $, respectively. In Fig. \ref{F2} and \ref{F3}, we plot the
tangential even/odd wave functions i.e., $Y_{k}^{\left( \pm \right) }\left(
z\right) $ and the corresponding probability densities i.e., $\left\vert
Y_{k}^{\left( \pm \right) }\left( z\right) \right\vert ^{2}$ in terms of $z$
for the first, second, third, and fourth quantized length-to-radius ratio for
each parity i.e., even and odd.

Our calculations show that with the $\frac{L}{R}=\lambda _{k}^{\left( \pm
\right) }$ the wave function of the particle with zero local energy makes
constructive interference with itself such that the stationary wave function
is nonzero and normalizable. On the other hand when $\frac{L}{R}\neq \lambda
_{k}^{\left( \pm \right) }$ the wave function makes a kind of destructive
interference with itself such that its wave function eventually vanishes.
Such particles cannot exist with zero local energy and must jump on any
other possible energy level corresponding to a wave function that satisfies
all boundary conditions. We assume hypothetically that there exists a
mechanism upon which we can measure the local energy of a quantum particle
confined on a spherocylindrical capsule. Since the zero energy corresponds
to a quantized $\frac{L}{R}$, a nonzero wave function for such a particle
implies the size of the capsule and vice versa. Therefore, we may have
created an indirect method to measure the size of nanotubes as well as the
energy of the quantum particles on those nanotubes.

\section{Conclusion}

In this research, we studied a spinless nonrelativistic quantum particle on
a spherocylindrical capsule. Employing Costa's formalism \cite{Costa}, we
solved the corresponding Schr\"{o}dinger equation by separating it into a
normal/global component multiplied by a tangent/local component. Assuming a
homogeneous and very thin surface, the normal or global component simply
became a one-dimensional infinite well problem with exact eigenfunctions and
eigenenergies. The tangent or local component, however, needed to be treated
in three different regions on the surface of the capsule. After we assumed
the tangent or local energy of the particle to be zero, we were able to
solve the 2-dimensional Schr\"{o}dinger equation on the three different
regions on the surface. Moreover, upon imposing the boundary conditions
consisting of the continuity of the wave function as well as its first
derivative, we obtained a relation between the geometric structure of the
capsule and the existence of a nonzero wave function. In summary, we can
state the results as follows: While a spinless nonrelativistic particle with
energy $E_{n}=\frac{n^{2}\pi ^{2}\hbar ^{2}}{2M\varepsilon ^{2}},$ $%
n=1,2,3,...$ can live on a spherical surface of radius $R$ and thickness $%
\varepsilon $ with a uniform probability distribution everywhere on the
surface, the same particle cannot survive on the surface of a
spherocylindrical capsule of the same radius $R$ for the hemispheres, length 
$L$ for the cylindrical body and thickness $\varepsilon $ unless $\frac{L}{R}
$ is equal to an even multiplication of $\pi $. Precisely, we found $\frac{L%
}{R}=2\left( 2k+1\right) \pi ,4k\pi $ where $k=0,1,2,...$. Physically it can
be understood in the context of the constructive or destructive interference
of the wave function of the particle with itself for $\frac{L}{R}=2\left(
2k+1\right) \pi ,4k\pi $ or $\frac{L}{R}\neq 2\left( 2k+1\right) \pi ,4k\pi ,
$ respectively. \ On one hand, the mechanism of having a non-zero wave
function with a given energy on the surface of the capsule may be used to
measure precisely the size of such nanotubes. On the other hand upon knowing
the size of the capsule and a nonzero wave function on the surface
hypothetically one may find the energy of such a particle.

\end{document}